\documentstyle[preprint,revtex,eqsecnum]{aps}
\tightenlines
\begin{document}
\preprint{McGill/92-17.}
\vskip 6 pt
\preprint{April 16, 1992.}
\draft
\begin{title}
{\bf
Virtual bremsstrahlung from pions and quarks\\
in thermalized hadronic matter}
\end{title}
\author{K. Haglin, C. Gale and V. Emel'yanov\cite{VE}}
\begin{instit}
Physics Dept., McGill University, Montr\'eal, P.Q., H3A 2T8, Canada
\end{instit}
\vskip 1.0 true in
\centerline{\bf Abstract}
\vskip -.25 true in
\begin{abstract}
A soft photon approximation is used to calculate the rates of
lepton pair production through virtual bremsstrahlung from both
pions and quarks.
Standard assumptions about the evolution of a nuclear system
under collision allow pion and quark driven total production to be
calculated.  Comparisons are made with Dalitz decay of light mesons.
These mechanisms are expected to be significant
contributors to the soft dilepton mass spectra one might observe
in heavy ion collisions at RHIC and LHC energies.
\end{abstract}

\narrowtext

Thermal photons and dileptons were among the first tools proposed for the
observation of thermalized nuclear matter \cite{ef76}. Leptons and photons
have small cross sections with the nuclear medium, so they will escape
without much further interaction from a region of dense
hadronic matter.  This allows them, at least in principle,
to carry information about
the temperature of the primordial state, which
may be a thermalized hadronic gas or a quark-gluon plasma (QGP).  We
shall consider here dilepton production from each,
concentrating in particular on electron-positron pairs.  The
dilepton invariant mass naturally subdivides into four distinct domains.
For masses $M$ above the $J/\psi$ peak, the spectrum is
expected to be dominated by Drell-Yan production \cite{es78}.
The spectrum around 1 GeV is dominated by $\rho, \omega$ and $\phi$ decays.
Dileptons with invariant masses in the 1.5--3.0 GeV region seem to
be mostly thermal \cite{es80}.  Competing sources in this mass region are
$D \bar D$ decays \cite{as89}, Drell-Yan processes, and pre-equilibrium
emission \cite{{ke90},{ma91}}.
The last region of
invariant mass ($M < m_{\rho}$) is a region of {\em soft lepton}
production \cite{es80}.
It is expected that below the $\rho$ mass a non-perturbative
QGP-induced dilepton radiation may appear.  There are many sources
of dileptons with invariant masses less than $m_{\rho}$.  For instance,
there are $\pi^{0}$ and $\eta$ Dalitz decays, $\pi^{+} \pi^{-}$ annihilation
from a thermalized pion gas, virtual bremsstrahlung with subsequent
decay from pions and quarks, and finally, quark-antiquark annihilation
in a QGP.
Recent calculations \cite{jc91} have shown that for masses near
$2m_{\pi}$, the contributions from $\pi^{0}$ and
$\eta$ Dalitz decays are two or three orders of magnitude above the
$q \bar q$ annihilation spectrum.  However, it has been shown for
zero $p_{T}$ dileptons having small invariant masses, that the
perturbative corrections can be several orders of magnitude
larger than the Born term \cite{eb90}. So, we consider this
$q \bar q$ annihilation spectrum as a lower limit only.

The contribution to the invariant mass spectrum from pion scattering
with virtual bremsstrahlung shown in Ref.~\cite{jc91}
is the same order as the $\pi^{0}$ and $\eta$ Dalitz decays
in the region $M \leq 2m_{\pi}$.  However, the authors considered
only the processes $\pi^{\pm} \pi^{0} \rightarrow \pi^{\pm}
\pi^{0}\gamma^{*}$.  Our purpose here is to supplement their results by
generalizing the formulae to include other pion-pion reactions, such as
$\pi^{\pm} \pi^{\mp} \rightarrow \pi^{\pm} \pi^{\mp}\gamma^{*}$.
Contributions from
$\pi^{\pm} \pi^{\pm} \rightarrow \pi^{\pm} \pi^{\pm}\gamma^{*}$ are
ignored because one expects significant cancellation from like-charge
interference \cite{klh91}.  Note the important difference
in the ordering of the charges.    Furthermore, we use these generalizations
along with lattice gauge theory data on screening to estimate
pair production from quark-quark and quark-gluon scattering processes
within a soft photon approximation \cite{rr76}.

We imagine the following picture for $e^{+}e^{-}$ production through
virtual bremsstrahlung in thermalized hadronic matter.  If the
initial temperature of the matter is $T_{i} > T_{c} = 200$ MeV, then
virtual bremsstrahlung from quarks occurs.
As the system evolves in time it cools.  If the
{\em QGP-hadron} phase transition is first order at $T=T_{c}$ the
mixed phase may be realized.  In this state there will be
dilepton emission from both quarks and pions.  Then, for
temperatures lower than $T_{c}$, say  $T_{c} > T > T_{f} \simeq 140$ MeV,
one need only consider radiation from pions.  Let us now estimate the rates
of $e^{+}e^{-}$ production in the quark and pion phases in the pursuit of a
time-evolution integrated pair production mass spectrum.

In the reaction $\pi \pi \rightarrow \pi \pi$ one expects
radiation of real and virtual photons as depicted in Fig.~\ref{qaqbqcqd},
with the external lines being pions and the shaded region representing
the strong interaction.
In the soft photon approximation \cite{cg87}, to which we restrict our
discussion, only the initial and final states radiate virtual quanta.
In this approximation, the
differential cross section for $e^{+}e^{-}$ pair production is
\begin{eqnarray}
{d \sigma^{e^{+}e^{-}}_{\pi \pi} \over dM^{2}} & = &
{\alpha^{2} \over 8\pi^{4}} {1\over M^{2}} \int
\left|\epsilon \cdot J \right|^{2}
{d \sigma_{\pi \pi} \over dt} \nonumber\\
& \ & \times \delta \left(M^{2} - (p_{+}
+ p_{-})^{2}\right)  {d^{3}p_{+}d^{3}p_{-} \over E_{+}E_{-}} dt
\label{eq:diffcross}
\end{eqnarray}
where $t$ is the four-momentum transfer in the $\pi \pi$ collision,
$M$ is the invariant mass of the $e^{+}e^{-}$ pair, $p_{+}$ and
$p_{-}$ are the momenta of the positron and electron respectively,
and
\begin{equation}
\left| \epsilon \cdot J \right|^{2} = \sum\limits_{\delta}
\epsilon_{\delta} \cdot J \ \epsilon_{\delta} \cdot J.
\label{eq:epsdotj}
\end{equation}
The sum is taken over all photon polarizations, and the current $J$ in the
reaction $\pi_{a} \pi_{b} \rightarrow \pi_{c} \pi_{d}$ is
\begin{equation}
J^{\mu} = - \widehat Q_{a} {p_{a}^{\mu} \over p_{a} \cdot q}
 - \widehat Q_{b} {p_{b}^{\mu} \over p_{b} \cdot q} + \widehat Q_{c}
{p_{c}^{\mu}
\over p_{c} \cdot q} + \widehat Q_{d} {p_{d}^{\mu} \over p_{d} \cdot q},
\label{eq:current}
\end{equation}
where $\widehat Q_{i} \equiv Q_{\pi_{i}}/|e|$ and $Q_{\pi_{i}}$ is
the charge of the {\em i \/}th pion.  When one of the
incoming pions is electrically neutral, Eq.~(\ref{eq:epsdotj}) can be
approximated for $|t| < 4m_{\pi}^{2}$
by \cite{cg87}
\begin{equation}
\left| \epsilon \cdot J\right|^{2} = {2\over 3} {1\over q_{0}^{2}}
\left( {-t\over m_{\pi}^{2}}\right).
\label{eq:pppzedj}
\end{equation}
The square of the photon's energy, $q_{0}^{2}$, is then replaced by
the symmetrized combination $E(E^2-M^2)^{1/2}$, with $E = E_{+} +
E_{-} = |\vec p_{+}| + | \vec p_{-} |$.   Within the same region of
validity, namely, $|t| <
4m_{\pi}^{2}$, we calculate $|\epsilon \cdot J|^{2}$ for all possible
charge combinations in $\pi_{a}\pi_{b} \rightarrow \pi_{a}\pi_{b}$
($\pi^{\pm}\pi^{0} \rightarrow \pi^{\pm}\pi^{0}$ and
$\pi^{\pm}\pi^{\mp} \rightarrow \pi^{\pm}\pi^{\mp}$) to be
\begin{equation}
\left| \epsilon \cdot J\right|^{2} = {2 \over 3} {1\over q_{0}^{2}}
\left( {-t\over m_{\pi}^{2}}\right) \left[ ( \widehat Q_{a}^{2}
+ \widehat Q_{b}^{2}) -{3\over 2} \widehat Q_{a} \widehat Q_{b}f(s) \right],
\label{eq:pppmedj}
\end{equation}
where
\begin{eqnarray}
f(s) & = & {s \over 2(s-4m_{\pi}^{2})} - {s-4m_{\pi}^{2}\over 2s}
\nonumber\\
& \ & -{m_{\pi}^{2} \over s} \left\lbrace
{2 \sqrt{s} \over \sqrt{s-4m_{\pi}^{2}}}
 + \left({\sqrt{s} \over \sqrt{s-4m_{\pi}^{2}}}\right)^{3} \right.
\nonumber\\
& \ &\left. + {\sqrt{s-4m_{\pi}^{2}}\over \sqrt{s}} \right\rbrace
\ln {\left| {\sqrt{s} + \sqrt{s-4m_{\pi}^{2}} \over
\sqrt{s} - \sqrt{s-4m_{\pi}^{2}}} \right|}.
\label{eq:fofs}
\end{eqnarray}
The interference function $f(s)$ behaves in the following manner.
In the limit $\sqrt{s} \rightarrow 2m_{\pi}$, $f(s) \rightarrow -2/3$ and
as $\sqrt{s} \rightarrow \infty$, $f(s) \rightarrow 0$.  In the small
$\sqrt{s}$ region ($\sqrt{s} \rightarrow 2m_{\pi}$) the factor
$|\epsilon \cdot J|^{2}_{\pi^{\pm}\pi^{\mp}} =
|\epsilon \cdot J|^{2}_{\pi^{\pm}\pi^{0}}$, but for $\sqrt{s} \rightarrow
\infty, \ |\epsilon \cdot J|^{2}_{\pi^{\pm}\pi^{\mp}} =
2|\epsilon \cdot J|^{2}_{\pi^{\pm}\pi^{0}}$.

The momentum integration in Eq.~(\ref{eq:diffcross}) gives
\begin{eqnarray}
{d\sigma^{e^{+}e^{-}}_{\pi \pi} \over dM^{2}} & = & {\alpha^{2}
\over 3\pi^{2}}
{\bar \sigma (s) \over M^{2}} \ln \left[ {\sqrt{s} - 2m_{\pi} \over
M}\right] \nonumber\\
& \ & \times\left\lbrace (\widehat Q_{a}^{2} + \widehat Q_{b}^{2}) -
{3 \over 2}
\widehat Q_{a} \widehat Q_{b} f(s) \right\rbrace,
\label{eq:sig}
\end{eqnarray}
where $\bar \sigma (s)$ is the momentum transfer weighted
cross section.  If $d\sigma/dt$ is a symmetric function of $t$ and
$u$, then
\begin{equation}
\bar \sigma(s) = 2\sigma_{el}(s) \left[{s\over 4m_{\pi}^{2}} - 1 \right].
\label{eq:barsig}
\end{equation}
The elastic pion-pion cross section is parametrized the same as
in Ref.~\cite{jc91}.
For $\sqrt{s} \leq$ 0.6 GeV the chiral model expression is adopted,
the $\rho$-resonance form of $\sigma_{el}$ is used for $0.6< \sqrt{s} \leq$
1.5 GeV and finally, a constant value of $5 \ mb$ is taken for
$\sqrt{s} >$ 1.5 GeV.
The $e^{+}e^{-}$ production rate through virtual bremsstrahlung
from pions is \cite{cg87}
\begin{eqnarray}
{dN \over d^{4}x dM^{2}} & = & {T^{6}\over (2\pi)^{4}}
\int\limits_{z_{\rm min}/T}^{\infty} dz z^{2} \nonumber\\
& \ & \times\left( z^{2} -
{4m_{\pi}^{2}\over T^{2}} \right) {\cal K}_{1}(z)
{d\sigma_{\pi \pi}^{e^{+}e^{-}} \over dM^{2}}
\label{eq:pionrate}
\end{eqnarray}
where $z_{\rm min} = 2m_{\pi} + M$, $d^{4}x$ is an element of
four-volume of the hadronic matter, and ${\cal K}_{1}$ is the modified
Bessel function.  In Fig.~\ref{chir} we show the
rate of $e^{+}e^{-}$ production from the sum of
$\pi^{\pm} \pi^{0}$ and $\pi^{\pm}\pi^{\mp}$
scattering at a temperature $T=200$ MeV.
Also in Fig.~\ref{chir} we compare our calculations with the rate of
production using the chiral model with quartic pion interactions
${\cal L}_{I} = \lambda (\vec \pi \cdot \vec \pi )^{2}/4$
($\lambda=1.4$ from $\pi \pi$ scattering lengths) \cite{aw91}.
Our results reproduce the rate from the chiral model
in the low mass region $M < 100$ MeV only if $\sqrt{s}$ is restricted
to the non-resonance region, say $\sqrt{s} \leq 0.48$ GeV.
Electron-positron pairs with invariant masses larger than $\sim 100$ MeV
are attributed to $\sqrt{s}$ in the resonance region and beyond.

The soft photon approximation has the advantage that
the electromagnetic and strong interaction components of the
matrix elements disentangle
leaving separate multiplicative factors (Low's theorem).
We utilize this property again to simplify the
contributions to the $e^{+}e^{-}$ spectrum from quark-quark and quark-gluon
scattering.  The same collection of diagrams from Fig.~\ref{qaqbqcqd} are
considered with the charged external lines being $u$ or $d$ quarks
(or antiquarks) and neutral external lines being gluons.  In total
we include six quark-quark (or antiquark) and four quark (or antiquark)-gluon
diagrams.  The electromagnetic {\em amplitude} is obtained from
Eq.~(\ref{eq:pppmedj}) by replacing the pion mass with the quark
mass and by setting $\widehat Q_{a}$ and $\widehat Q_{b}$ to the
appropriate quark
electric charges.  The strong
interaction cross sections ${d \sigma_{q g}/
dt}$ and ${d \sigma_{qq} / dt}$ are
well known in the perturbative vacuum at the one-gluon-exchange
level to be \cite{jl82}
\begin{equation}
{d \sigma_{a b} \over dt} = {C_{ab} 2\pi \alpha_{s}^{2} \over t^2},
\label{dsigqq}
\end{equation}
where
\begin{equation}
C_{ab} = \left\lbrace {1 {\ (qg\rightarrow \ qg)} \atop {4 \over 9}
{\ (qq\ \rightarrow \ qq)}} \right. .
\end{equation}
For hot hadronic matter this is clearly inadequate.
At finite temperatures, static color-electric fields
are shielded by quarks and gluons, and color-magnetic
fields are shielded by non-perturbative effects \cite{pd85}.  To include
such effects in a finite-temperature evaluation of
diagrams like the one in Fig.~\ref{qaqbqcqd},
we use the {\em color-electric mass} $m_{E}^{2} = 6\pi \alpha_{s}(t)T^2$,
and the lattice gauge result for the {\em color-magnetic mass}
$m_{M}^{2} = 25 \alpha_{s}^{2}(t) T^2$ \cite{td82} of the exchanged
gluon.  Then for massless quarks whose three-momenta are initially
anticollinear in the frame fixed by the finite-temperature medium, we
have \cite{pd85}
\begin{equation}
{d\sigma_{ab} \over dt} = C_{ab} {\pi \over 2} {\alpha_{s}^{2}(t)
(2t - m_{E}^{2} - m_{M}^{2})^{2} \over (t-m_{E}^{2})^2
(t-m_{M}^{2})^{2}}.
\label{dsabdt}
\end{equation}
We check the validity of using massless quarks by assuming
$\alpha_{E}(t) \simeq \alpha_{M}(t)$ and introducing finite
quark masses.  The dilepton production rates change very little,
so to this level of accuracy, Eq.~(\ref{dsabdt}) is acceptable.
Note that in the limit $m_{E},m_{M} \rightarrow 0$ (free space)
the resulting cross section reproduces
Eq.~(\ref{dsigqq}).  Realistically, $\alpha_{s}$ is also temperature
dependent owing to many-body effects.  Since we make small $|t|$
approximations throughout, we ignore the momentum dependence and
adopt the lattice gauge theory renormalization group
result for the temperature dependence \cite{fk88}
\begin{equation}
\alpha_{s}(T) = {6\pi \over (33-2N_{f})\ln(aT/T_{c})},
\end{equation}
with $a \sim 8$.

Having the strong interaction cross sections, we can use
Eq.~(\ref{eq:pionrate}) to calculate the rate of $e^{+}e^{-}$
production through virtual bremsstrahlung in $qq \rightarrow qq$
scattering by replacing the pion mass with the quark mass and
multiplying by the appropriate spin degeneracy factor 4.  For
$qg \rightarrow qg$ scattering we must use
\begin{eqnarray}
{dN \over d^{4}x dM^{2}} & = & {T^{6}\over (2\pi)^{4}}
\int\limits_{z_{\rm min}/T}^{\infty} dz \nonumber\\
& \ & \times\left( z^{2} -
{m_{q}^{2}\over T^{2}} \right)^{2} {\cal K}_{1}(z)
{d\sigma_{qg}^{e^{+}e^{-}} \over dM^{2}},
\label{eq:quarkrate}
\end{eqnarray}
where this time $z_{\rm min} = m_{q} + M$ and the spin degeneracy
to multiply Eq.~(\ref{eq:quarkrate}) is 6.
Kinematics also changes the argument of the logarithm in Eq.~(\ref{eq:sig})
to $(\sqrt{s} - m_{q})/M$.

A choice must be made for the quark mass to be used in
Eqs.~(\ref{eq:pppmedj}--\ref{eq:pionrate}) and ~(\ref{eq:quarkrate}).
Chiral invariance is
restored for temperatures near the phase transition, so
{\em constituent} quarks become {\em current} quarks with masses
$m_{q} \simeq 5$ MeV. On the another hand, in a QGP the fermion
mass induced by many-body effects is \cite{es78,es80}
$m_{q} = (2\pi \alpha_{s}/3)^{1/2} T$, which is much greater
than 5 MeV.  Since the quark-mass dependence in the rate is roughly
$1/m_{q}^{2}$ due to Eq.~(\ref{eq:barsig}) and the temperature
dependence is roughly $T^{6}$ as in Eq.~(\ref{eq:pionrate}), an
upper limit on the rate is obtained using $m_{q}= 5$ MeV at $T=300$ MeV.
Similarly, a lower limit on the rate is obtained by using
$m_{q}= 300$ MeV at the phase transition temperature.
We show in Fig.~\ref{rat} these lower and upper limits on the
quark-driven rate, the rate obtained using the induced quark mass
at two different values of temperature, and
finally, the pion-driven contribution at $T_{c}$.  From this one
may conclude that the pion gas is slightly more luminous than our
{\em medium} quark gas at $T_{c}$, for the range of invariant masses
we consider.  The rate of production from the {\em current} quarks
is much larger than the pion rate.

Total $e^{+}e^{-}$ production from competing quark and pion processes
is obtained only after integrating the rate from
Eq.~(\ref{eq:pionrate}) over the space-time history
of the nuclear system.  Using the standard 1+1 dimensional expansion
in a Bjorken picture \cite{jb83,kk86}, we reproduce the $\pi^{\pm}\pi^{0}$
results of Ref.~\cite{jc91} from mixed plus cooling phases using
their values for the relevant temperatures.  By including the
$\pi^{\pm}\pi^{\mp}$ reactions we find a 20\% enhancement
over their result.  The enhancement is smaller than one would
expect, but in our case the cross section is
$\sigma_{el} \sim \sigma^{\pi^{+}\pi^{0}} +
\sigma^{\pi^{-}\pi^{0}} + \sigma^{\pi^{+}\pi^{-}}$
with an equal weighting 1/3 of each channel, which is
different from Ref.~\cite{jc91} where
$\sigma_{el} \sim \sigma^{\pi^{+}\pi^{0}} +
\sigma^{\pi^{-}\pi^{0}}$ and therefore, the weighting is 1/2 for
both channels.  We go on to choose temperatures
$T_{i}=300$ MeV, $T_{c}=200$ MeV, and $T_{f}=140$ MeV in order to
investigate other conditions likely to be found in heavy ion collisions
at RHIC and LHC energies.  In Fig.~\ref{spec} we present the
resulting invariant mass spectra through virtual bremsstrahlung from
quarks and pions separately.  The quark result is the sum of contributions
from the cooling ($T_{i} > T > T_{c}$) plus mixed ($T = T_{c}$) phases and
the pion result is the sum of contributions from the mixed ($T = T_{c}$)
plus cooling ($T_{c} > T > T_{f}$) phases.   The pion cooling-phase
contribution is the larger contributor to its final spectrum; whereas
the mixed-phase dominates the resulting quark spectrum.
For a more complete comparison with other competing processes,
we show also the results from $\pi^{0}$
and $\eta$ Dalitz decay as well as
$q\bar q \rightarrow \gamma^{*} \rightarrow e^{+}e^{-}$ annihilation
in the Born approximation.
Using our values of $T_{i}, T_{c}$ and $T_{f}$, the {\em medium} quark
contribution (from cooling plus mixed phases)
and the {\em constituent} quark contribution (from the mixed phase only)
are both lower than the $\pi^{0}$ Dalitz,
the $\eta$ Dalitz, and well below the pion contribution.
On the other hand, the {\em current} quark contribution
is larger than $\pi^{0}$ or $\eta$ Dalitz; it is even larger
than the resulting pion contribution by a factor of 10 or more.
However, we believe the true quark contribution will
be nearer to the {\em medium} quark result.  This being true,
virtual bremsstrahlung from pions will be the largest source of
low-mass $e^{+}e^{-}$ pairs.  This points out the need for more
elaborate calculations, perhaps based on nonequilibrium approaches.

Assuming only soft virtual photons greatly simplified our task
by eliminating the need to exactly evaluate the many Feynman diagrams
that contribute to $e^{+}e^{-}$ production.   But the validity of
this popular approach may be questioned.
For instance, the restriction that the momentum
transfer $|t|$ be less than $4m_{\pi}^{2}$ for pions and less
than $4m_{q}^{2}$ for the quarks is arguably too restrictive.
Also, many-body effects introduce temperature dependences
into the meson masses, widths and coupling constants and therefore,
into the $\pi \pi$ elastic cross section likely to be quite strong
near $T_{c}$.  Detailed discussions of these effects will be presented in
a future paper \cite{cg92}.
\acknowledgements

One of us (V.E.) is indebted to Profs. S. K. Mark and P. Depommier for
helping organize his stay at McGill University.  This work was supported
by the National Science and Engineering Research Council of Canada
and by the FCAR fund of the Qu\'ebec
Government.

\figure{Lepton pair production through virtual bremsstrahlung in
scattering of particles $a \ b \rightarrow c \ d$.  The
particles might be charged or uncharged pions, quarks or antiquarks
or gluons. The shaded region indicates a strong interaction.\label{qaqbqcqd}}
\figure{Chiral model predictions for the $e^{+}e^{-}$ production rate as
compared with our calculation.  The dashed lines are our estimations
with varying values for the upper limit on the energy and the solid
line is from the chiral model.\label{chir}}
\figure{Mass dependence of quark-driven and pion-driven production
rates at temperatures $T=200$ MeV and $T=300$ MeV.\label{rat}}
\figure{Dilepton mass spectra for LHC energies:  The solid curve
is the total contribution from pion scattering processes, the dot-dashed
curves result from quark-quark and quark-gluon scattering with different
quark masses, the long-dashed curve is the result of $q \bar q$ annihilation
in a Born approximation, and finally, the short-dashed curves are
Dalitz decay spectra.
\label{spec}}

\end{document}